\def\year{2020}\relax
\documentclass[letterpaper]{article} 
\usepackage{aaai20}  
\usepackage{times}  
\usepackage{helvet} 
\usepackage{courier}  
\usepackage[hyphens]{url}  
\usepackage{graphicx} 
\usepackage{graphicx}  
\usepackage{algorithm}
\usepackage[noend]{algpseudocode}

\usepackage{amssymb}
\usepackage{microtype}
\usepackage{subcaption}

\usepackage{amsfonts}
\usepackage{amsthm}
\usepackage{amsmath}

\DeclareMathOperator*{\argmax}{arg\,max}

\newcommand{\fig}[1]{Figure~\ref{fig:#1}}

\newcommand{\tab}[1]{Table~\ref{tab:#1}}
\newcommand{\alg}[1]{Algorithm~\ref{alg:#1}}
\newcommand{\eq}[1]{(\ref{eq:#1})}

\usepackage{enumitem}
\usepackage{booktabs}

\newsavebox{\tempfig}

\title{Relevance Proximity Graphs for Fast Relevance Retrieval}
\author{Stanislav Morozov \\ Yandex, \\ Lomonosov Moscow State University \\ \tt\small  stanis-morozov@yandex.ru
\And
Artem Babenko \\ Yandex, \\ National Research University \\Higher School of Economics \\ \tt\small  artem.babenko@phystech.edu}
 
\begin{document}

\maketitle

\begin{abstract}
In plenty of machine learning applications, the most relevant items for a particular query should be efficiently extracted, while the relevance function is based on a highly-nonlinear model, e.g., DNNs or GBDTs. Due to the high computational complexity of such models, exhaustive search is infeasible even for medium-scale problems. To address this issue, we introduce \textbf{Relevance Proximity Graphs~(RPG)}: an efficient non-exhaustive approach that provides a high-quality approximate solution for maximal relevance retrieval. Namely, we extend the recent similarity graphs framework to the setting, when there is no similarity measure defined on item pairs, which is a common practical use-case. By design, our approach directly maximizes off-the-shelf relevance functions and does not require any proxy auxiliary models. Via extensive experiments, we show that the developed method provides excellent retrieval accuracy while requiring only a few model computations, outperforming indirect models. We open-source our implementation as well as two large-scale datasets to support further research on relevance retrieval.
\end{abstract}

\section{Introduction}
\label{sect:intro}
Retrieval of most relevant items for a particular query is a key ingredient of a wide range of machine learning applications, e.g., recommender services, retrieval-based chatbot systems, web search engines. In these applications, item relevance for a particular query is usually predicted by pretrained models, such as deep neural networks (DNNs) or gradient boosted decision trees (GBDTs). Typically, relevance models get a query-item pair as input and output the relevance of the item to the query. In the most general case, queries and items are described by different sets of features and belong to different spaces. For instance, in recommender systems, queries are typically described by user gender, age, usage history, while the item features mostly contain information describing the content. Let us denote the \textit{query} space by $Q$ and the \textit{item} space by $V$. Then the problem of maximal relevance retrieval can formally be stated as follows. Let us have a large finite set $S$ of items $S \subset V$, query $q \in Q$ and the function $f(q, v)$ that maps query-item pairs to relevance values:
\begin{equation}
     f: Q \times V \rightarrow \mathbb{R}
\end{equation}
For a given query $q \in Q$ we aim to find an item that maximizes the relevance function $f(q, v)$:
\begin{equation}
    \argmax\limits_{v \in S} f(q, v)
    \label{eq:problem}
\end{equation}
or, more generally to find top-$K$ items from $S$ that provide maximal relevance values. An important special case of problem \eq{problem} when $Q = V = \mathbb{R}^n$ and
\begin{equation}
    f(q,v) = -\|q - v\|^2
\end{equation}
is a well-known problem of nearest neighbor search, which was investigated by the machine learning community for decades \cite{bentley1975,LSH98,dasgupta2013randomized}.

However, current applications typically use more complex and highly-nonlinear relevance functions $f(\cdot, \cdot)$. For instance, many modern recommender services \cite{he2017neural} use the deep neural networks for relevance prediction, while chatbots \cite{chatbot18} and many other applications use GBDT models. Naive exhaustive search requires $|S|$ relevance function computations, which is not feasible for million-scale databases and computationally expensive models. In this paper, we propose a method that provides an approximate solution of high quality while computing $f$ only for a small fraction of $S$.

The proposed method expands the approach of similarity graphs, which was shown to provide exceptional performance for the nearest neighbor search problem  \cite{navarro2002searching,malkov2018efficient,NSG}. This approach organizes the set of items in a graph, where close items are connected by edges, and the search process is performed via a greedy exploration of this graph. In this paper, we extend similarity graphs to the setting, when there is no similarity measure defined on item pairs. Specifically, we describe each item by a vector of relevance values for a representative subset of queries, and experimentally show that the graph exploration can be successfully guided by the DNN/GBDT models. Below we refer to our method as \textbf{\textit{Relevance Proximity Graphs} (RPG)}.

The contributions of the paper can be summarized as follows:
\begin{enumerate}
    \setlength\itemsep{0em}
    \item We tackle a new problem of general relevance retrieval, which so far received little attention from the community.
    \item We extend the similarity graphs framework to the setting without a similarity measure defined in the item space.
    \item We open-source the implementation\footnote{\url{https://github.com/stanis-morozov/rpg}} of the proposed method as well as two million-scale datasets and state-of-the-art pretrained models for further research on general relevance retrieval.
\end{enumerate}

The rest of the paper is organized as follows: in Section~\ref{sect:related}, we briefly review prior work related to the proposed approach. Section~\ref{sect:theory} formally describes the construction and the usage of RPG. In Section~\ref{sect:experiments}, we perform an extensive evaluation of RPG on several datasets and several models and empirically confirm its practical usefulness. Section~\ref{sect:conclusion} concludes the paper.

\section{Related work}
\label{sect:related}
\textbf{Relevance retrieval problem.} Probably, the closest work to ours is  \cite{zhu2018learning}, which also notes that large-scale relevance retrieval is computationally infeasible for multi-layer DNN models. They tackle this problem by learning a hierarchical model with a specific structure, which organizes the set of items into a tree during the training stage. While this model allows non-exhaustive retrieval, their approach cannot be directly applied to existing models and requires training the model from scratch. Another work \cite{wang2011cascade} proposes a cascade scheme when cheap auxiliary models provide short-lists of promising candidates and expensive relevance models are used only for candidate reranking. While being efficient, such schemes can result in low recall if the capacity of auxiliary models is insufficient to extract precise candidates lists. In contrast, the proposed RPG approach directly maximizes relevance given by arbitrary off-the-shelf models. We confirm this claim for several GBDT and DNN models in the experimental section.

\textbf{Nearest neighbor search.} As mentioned above, problem \eq{problem} generalizes the well-known problem of nearest neighbor search (NNS). Overall, machine learning community includes three separate lines of research on NNS: locality-sensitive hashing (LSH) \cite{LSH98,andoni2008near}, partition trees \cite{bentley1975,Dasgupta2008,dasgupta2013randomized} and similarity graphs \cite{navarro2002searching}. While LSH-based and tree-based methods provide solid theoretical guarantees, the performance of graph-based methods was shown to be much higher \cite{malkov2018efficient}. The proposed approach is based on the similarity graph framework, hence we provide a brief review of its main ideas below.

For a set of items $S$, the directed similarity graph has a vertex corresponding to each of the items. Vertices $v_i$ and $v_j$ are connected by an edge if $v_j$ belongs to the set of $k$ nearest neighbors of $v_i$ in terms of similarity function $s$. The usage of similarity graphs for the NNS problem was initially proposed in the seminal work \cite{navarro2002searching}. This approach constructs the similarity graph and then performs the greedy walk on this graph on the retrieval stage. The search process starts from a random vertex and then on each step moves from the current vertex to a neighbor, that appears to be the closest to the query. The process terminates when we reach a local minimum, i.e., there are no adjacent vertices, closer to the query. Since  \cite{navarro2002searching} the general pipeline described above, was modified by a large set of additional heuristics \cite{malkov2018efficient,fu2016efanna,NSG}, outperforming LSH-based and tree-based methods.

The proposed approach for the relevance retrieval problem could be based on any state-of-the-art similarity graph. In this paper, we employ Hierarchical Navigable Small World (HNSW) \cite{malkov2018efficient} graphs implementation that is  publicly available. In HNSW, the search is performed in a semi-greedy fashion, via a variant of beam search \cite{Bisiani87}, as described in \alg{greedy_exploration} in detail. During the construction stage, HNSW builds the graph via incrementally adding vertices one-by-one. There is a parameter $M$, denoting for the maximum degree of vertices in the graph. For each vertex $v$ we perform the search described in \alg{greedy_exploration} and connect $v$ to the $M$ closest vertices that already exist in the graph. HNSW builds a nested hierarchy of graphs, where the lowest layer contains all vertices, and each higher layer contains only a subset of the vertices of the lower layer. The search is performed from top to bottom and the result from each upper layer is used as an entry point to the lower layer.

Beyond the NNS problem, similarity graphs were also shown to provide decent performance for more general similarity functions, e.g., inner product, Kullback-Leibler divergence, cosine distance, Itakura-Saito distance, and others \cite{boytsov2017efficient}.
In this paper, we show that this framework can be extended to work with the setting where there is no specified similarity measure between the item space elements and the relevance functions defined by state-of-the-art ML models, such as DNNs or GBDTs.


\section{Relevance Proximity Graphs (RPG)}
\label{sect:theory}
The key idea of our approach is to represent the set of items as a graph and to perform the search on this graph using the given relevance function. The retrieval stage remains unchanged, that is, we perform semi-greedy graph exploration, guided by the relevance function $f(\cdot, \cdot)$ (\alg{greedy_exploration}). As will be shown in the experiments, the state-of-the-art DNN and GBDT models successfully guide the graph exploration process given that the item set is organized in an appropriate graph.

However, the question ''How to construct an appropriate graph?'' becomes nontrivial as items and queries belong to different spaces. Moreover, in some scenarios, there is no similarity defined in the item space, hence the existing approaches to graph construction cannot be directly applied.
\begin{algorithm}
   \caption{Graph Exploration}
   \label{Greedy_search}
\begin{algorithmic}
   \State {\bfseries Input:} Graph $G$, relevance function $f(q,v)$, query $q$, entry vertex $v_{0}$, beam width $L$
   \State Candidate set $C = \{v_{0}\}$
   \State Set of visited vertices $V = \{v_{0}\}$
   \State List of the most relevant vertices $W = \{v_{0}\}$
   \While{$|C| > 0$}
   \State extract to $v_{curr}$ the most relevant element from $C$
   \State extract to $b$ the less relevant element from $W$
   \If{$f(q, v_{curr}) < f(q, b)$}
   \State \textbf{break}
   \EndIf
   \For{$v_{adj}$ adjacent to $v_{curr}$ in $G$}
   \If{$v_{adj} \not\in V$}
   \State $V = V \cup \{v_{adj}\}$
   \If{$f(q, v_{adj}) > f(q, b)$ or $|W| < L$}
   \State $C = C \cup \{v_{adj}\}$
   \State $W = W \cup \{v_{adj}\}$
   \If{$|W| > L$}
   \State erase the less relevant element from $W$
   \EndIf
   \EndIf
   \EndIf
   \EndFor
   \EndWhile
   \State \textbf{return} $W$
\end{algorithmic}
\label{alg:greedy_exploration}
\end{algorithm}

\subsection{Relevance-aware similarity in the item space}

For graph construction, we exploit the natural idea that two items $u$ and $v$ are similar if the corresponding functions $f(\cdot,u)$ and $f(\cdot,v)$ are ''close'', i.e. the items are both relevant or irrelevant for the same query.  
As a straightforward way to define the similarity between functions, we use $L_2$ distance over some measure $\mu$ defined on the query space (we put minus sign as similarity search is traditionally formulated as a maximization problem):
\begin{equation}
    \label{general_similarity}
    s(u, v) = -\Big( \int\limits_{Q} (f(q, u) - f(q, v))^2 d\mu(q) \Big)^{1/2}
\end{equation}
We choose the proper measure over the query space $\mu$ based on the following intuition. Let us have a probability space $(Q, \mathcal{F}, \mathbb{P})$ defined on the query space. For most applications, it is natural to force functions $f(q, u)$ and $f(q, v)$ to be closer in the regions where the density of the query distribution is high. Then, it corresponds to the following similarity function:
\begin{equation}
    s(u, v) = -\Big( \int\limits_{Q} (f(q, u) - f(q, v))^2 d\mathbb{P}(q) \Big)^{1/2}
\end{equation}
that is equivalent to the expectation over the probability space:
\begin{equation}
\label{precise_similarity}
    s(u, v) = -\Big( \mathop{\mathbb{E}}_{q\sim Q} (f(q, u) - f(q, v))^2 \Big)^{1/2}
\end{equation}
In practice, we use the Monte-Carlo estimate of this value. Let us have a random sample $X$ of size $d$ from the query distribution. We enumerate this random sample:
\begin{equation}
    X = \{ q^{(1)}, q^{(2)}, \dots, q^{(d)} \}
    \label{eq:sample}
\end{equation}
Then we define a vector $r_u$ corresponding to the item $u$ in the following way:
\begin{equation}
    r_u^i = f(q^{(i)}, u)
\end{equation}
We refer to the vector $r_u \in \mathbb{R}^d$ as a \textit{relevance vector} as it contains the relevance values for the item $u$ and queries in the sample $X$. Note that we choose the sample $X$ only once and it remains the same for all the items. Then the similarity between items $u$ and $v$ can be defined as:
\begin{equation}
\label{modelawaresimilarity}
    s(u, v) = -\| r_u - r_v \|
\end{equation}

Given this similarity measure, we can apply the existing graph construction method from \cite{malkov2018efficient}. Note, that for the fair evaluation, only hold-out training queries were used to obtain the relevance vectors in all the experiments, while the relevance retrieval accuracy was calculated for a separate set of test queries.

\subsection{RPG construction}
We summarize the graph construction scheme more formally. Let us have the item set $S \subset V$ and the train query set $\{q_1,\dots,q_N\}$. The main parameter of our scheme is a dimensionality of relevance vectors, which is denoted by $d$.

\begin{enumerate}
    \item Select $X$ --- $d$ queries from $\{q_1,\dots,q_N\}$, which will be used to construct the relevance vectors.
    \item Compute the relevance vectors for items from $S$: $\{ r_u^i\}_{u \in S, q^{(i)} \in X} = \{ f(q^{(i)}, u)\}_{u \in S, q^{(i)} \in X}$
    \item Build a similarity graph on $S$, using $L_2$ distance metric on the relevance vectors as a similarity measure, via HNSW method \cite{malkov2018efficient}.
\end{enumerate}

Given the graph, the maximal relevance retrieval is performed for a query $q$ via \alg{greedy_exploration}, using the model $f(\cdot, \cdot)$ to guide the graph traversal. As in \cite{malkov2018efficient}, for all queries, we use the same entry vertex $v_0$ in RPG, which corresponds to the item with id${=}0$.



\section{Experiments}
\label{sect:experiments}
\begin{figure*}
    \centering
    \begin{subfigure}[b]{0.49\textwidth}
        \includegraphics[width=\textwidth]{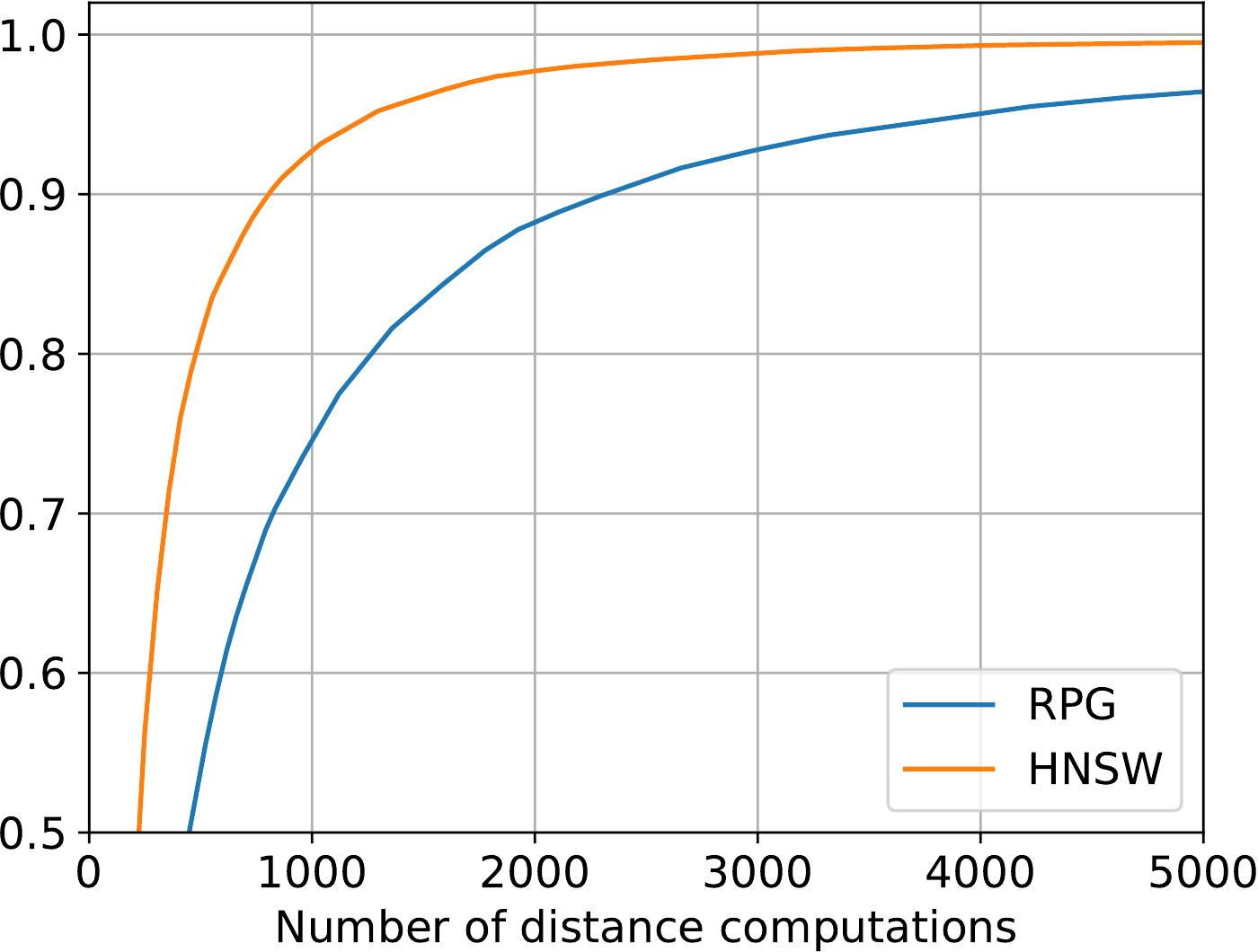}
        \caption{SIFT1M}
        \label{fig:RPGL2_SIFT}
    \end{subfigure}
    \begin{subfigure}[b]{0.49\textwidth}
        \includegraphics[width=\textwidth]{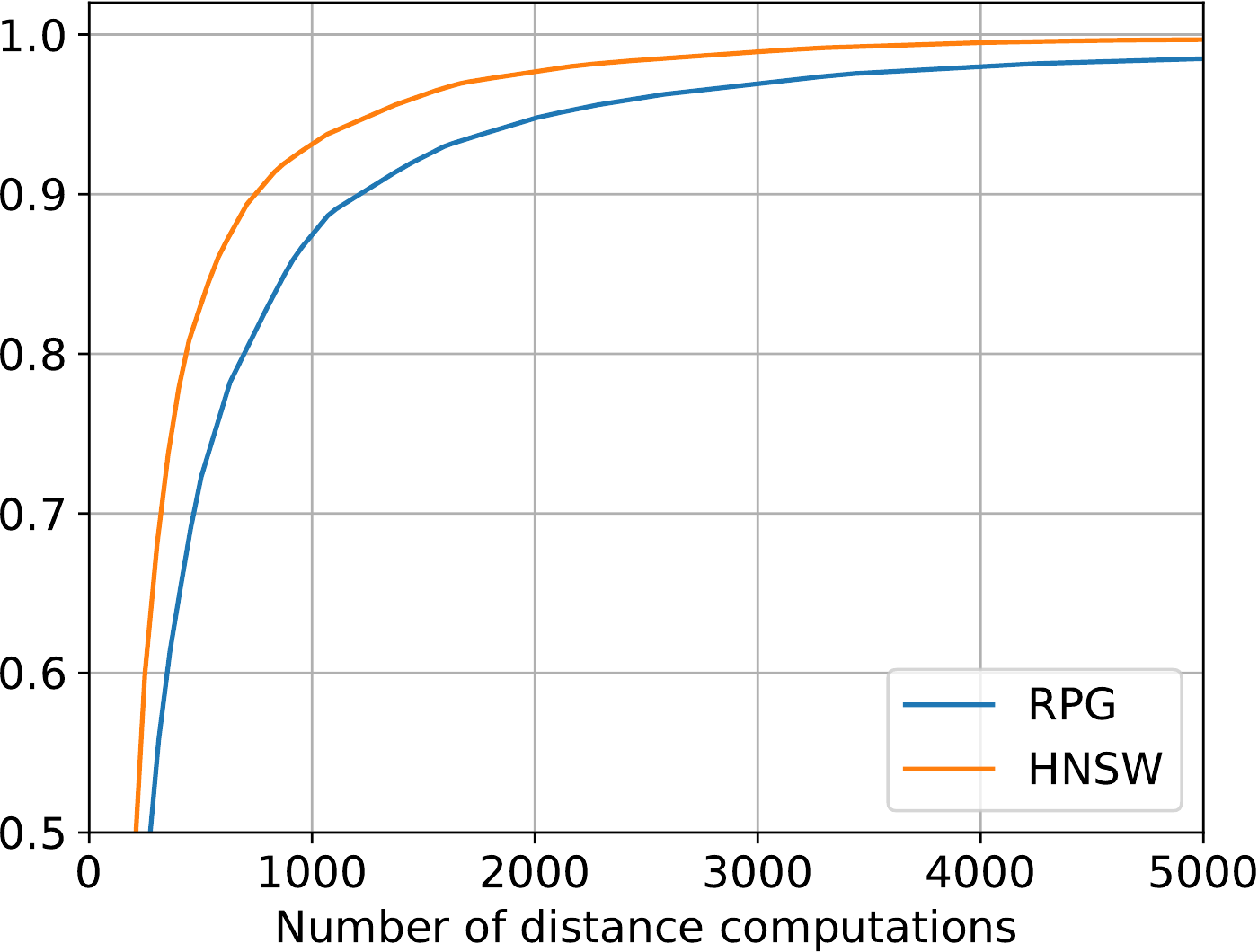}
        \caption{DEEP1M}
        \label{fig:RPGL2_DEEP}
    \end{subfigure}
    \caption{The comparison of RPG with HNSW for euclidean nearest neighbor benchmarks. For both datasets we retrieve top-$5$ items.}
\end{figure*}

In this section, we present the experimental evaluation of the proposed RPG approach for the top-K relevance retrieval problem on three real-world datasets. Our code is written in C++, and the implementation of similarity graphs is based on the open-source HNSW implementation\footnote{\url{https://github.com/yurymalkov/hnsw}}. In our experiments, we use two standard performance measures. Commonly used \textit{Recall} measure is defined as the rate of successfully found neighbors, averaged over a set of queries. The second is \textit{Average relevance} that is defined as the average of relevance values for the query and retrieved top-K items, averaged over the set of queries.

\subsection{Datasets}
We report experimental results obtained on three datasets described below. To the best of our knowledge, there are no publicly available large-scale benchmarks for relevance retrieval with highly-nonlinear models without the similarity measure between item space elements, therefore we collect and open-source two datasets. We expect that these datasets will be valuable for the community, given the abundance of relevance retrieval problem in applications.

\textbf{Collections dataset.} This dataset originates from a proprietary image recommendation service. Here we also sampled one million most-viewed images and $2,000$ random warm users. Each user-item pair is characterized by $138$ features, where $93$ of them are item features, $16$ are user features and $29$ are pairwise user-item features. Then we trained the state-of-the-art GBDT model \cite{Catboost} on these features, which we open-source along with the dataset.

\textbf{Video dataset.} This dataset originates from a proprietary video recommendation service. We sampled one million most-viewed videos and $2,000$ random warm users. Each user-item pair is characterized by $2,715$ features, where $562$ of them are item features, $2,080$ are user features and $73$ are pairwise user-item features. We trained the GBDT model \cite{Catboost} on these features, which we open-source as well.

\textbf{Pinterest.} We also evaluate our approach on the medium-scale \textit{Pinterest} dataset \cite{geng2015learning} with the deep neural network model for relevance prediction proposed in \cite{he2017neural}. This datasest contains $9,916$ items and $55,187$ queries without any feature representations, i.e. only a rating matrix is provided.


For all datasets we randomly selected $1,000$ users as train queries and $1,000$ users as test queries. We use the train queries for the relevance vector computation, and we average the evaluation measures over the test queries.

\subsection{Sanity check}
To demonstrate the reasonableness of the proposed scheme we evaluate RPG on two common nearest neighbor benchmarks \textit{SIFT1M} \cite{lowe2004distinctive} and \textit{DEEP1M} \cite{babenko2016efficient} with the euclidean distance between queries and items, that is $Q = V = \mathbb{R}^n$ and

\begin{equation}
    f(q, v) = -\| q - v \|^2
\end{equation}

On \fig{RPGL2_SIFT} and \fig{RPGL2_DEEP} we provide the results of the comparison of RPG with HNSW \cite{malkov2018efficient}. For both methods we use $M=8$ and $100$-dimensional relevance vectors for RPG. Indeed, the graphs constructed based on distances between relevance vectors (\ref{modelawaresimilarity}) are less accurate but still provide decent retrieval performance. In particular, it is sufficient to perform only a few thousand distance evaluations to achieve $0.9$ recall level.

We conjecture that the reason of the decent performance even with suboptimal graphs is that on the graph exploration stage the search process is "guided" by the correct similarity measure, which is negative $L_2$ distance between the original data vectors in this experiment. Furthermore, as we show in the experiments below, the graphs constructed on relevance vectors perform exceptionally well even when the relevance function $f(\cdot,\cdot)$ is based on highly-nonlinear DNN or GBDT models. 

\begin{figure}
    \centering
    \begin{minipage}[t]{0.45\textwidth}
        \centering
        \includegraphics[width=\textwidth]{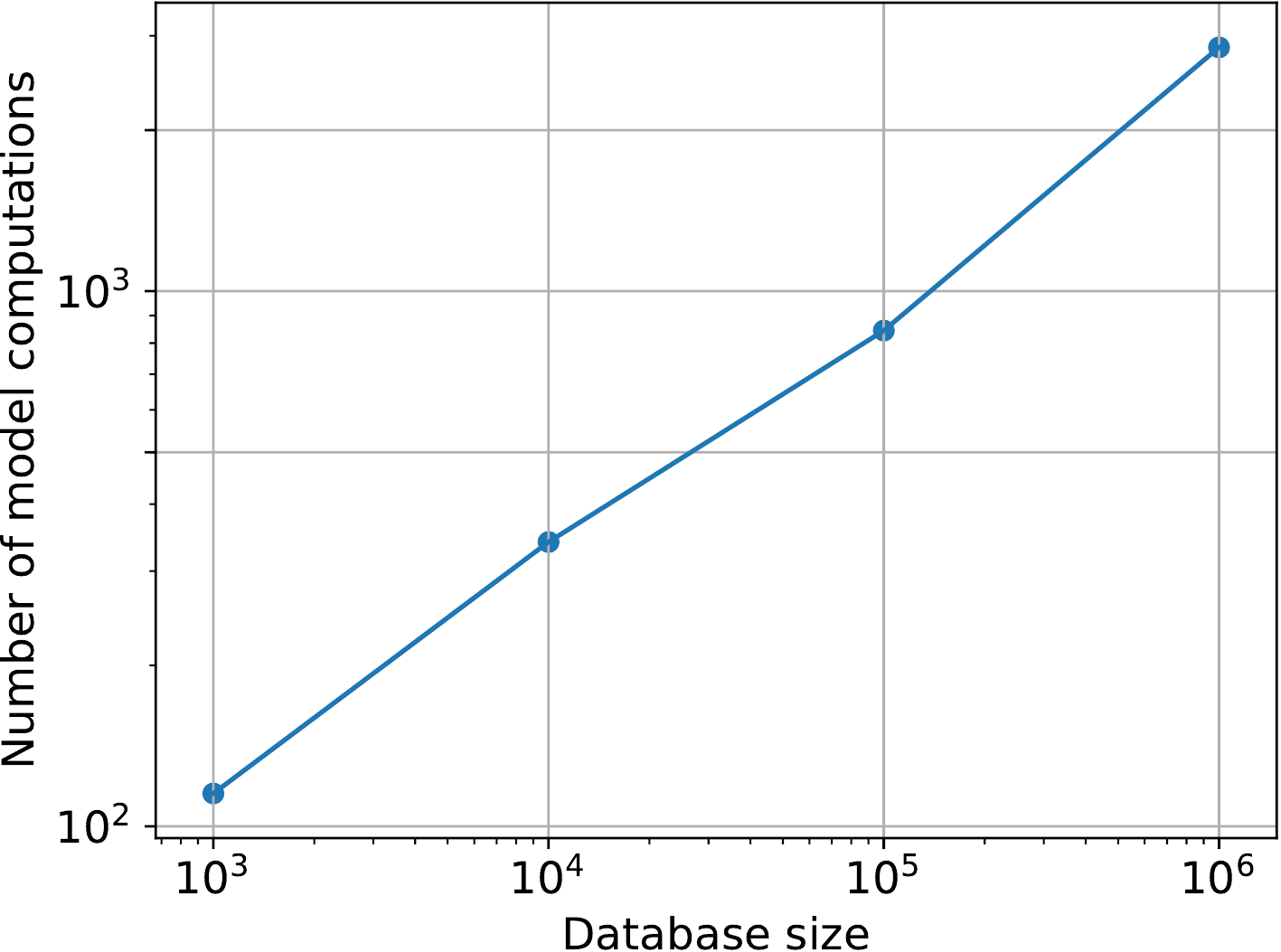}
        \caption{The dependence on the number of relevance function computations on database size to achieve $0.90$ \textit{Recall} on the \textit{Collections} dataset.}
        \label{fig:scalability}
    \end{minipage}
\end{figure}

\subsection{Ablation and preliminary experiments}

Now we investigate RPG performance with varying database sizes and different parameter values.

\begin{figure}
    \centering
    \begin{minipage}[t]{0.47\textwidth}
        \includegraphics[width=\textwidth]{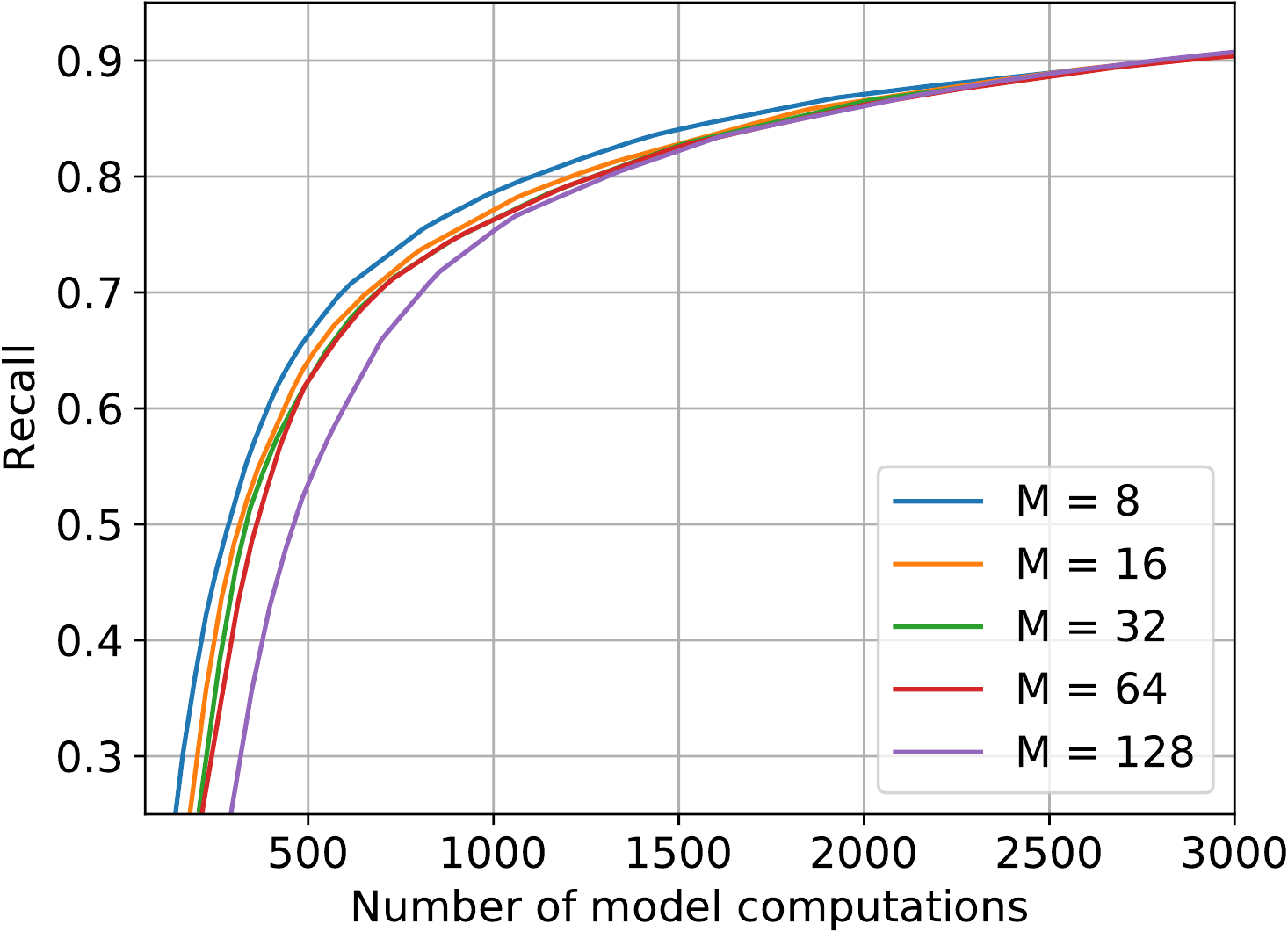}
        \caption{The RPG performance with the different vertex degree $M$ on the Collections dataset.}
        \label{fig:diff_M}
    \end{minipage}
\end{figure}

\textbf{RPG vertex degree.} First, we investigate the dependence of RPG performance on the vertex degree $M$. For the \textit{Collections} dataset, the recall-vs-complexity curves for the different $M$ values are shown in \fig{diff_M}. Here, the length of the relevance vectors is equal to $d{=}1000$ for all $M$ values. Surprisingly, \fig{diff_M} demonstrates that the best results are obtained for a quite small degree $M{=}8$, which is smaller than the typical vertex degrees in graphs for metric nearest neighbor search \cite{malkov2018efficient}. In all the experiments below we use $M{=}8$ for all datasets.

\begin{figure*}[!ht]
    \centering
    \begin{subfigure}[b]{0.3405\textwidth}
        \includegraphics[width=\textwidth]{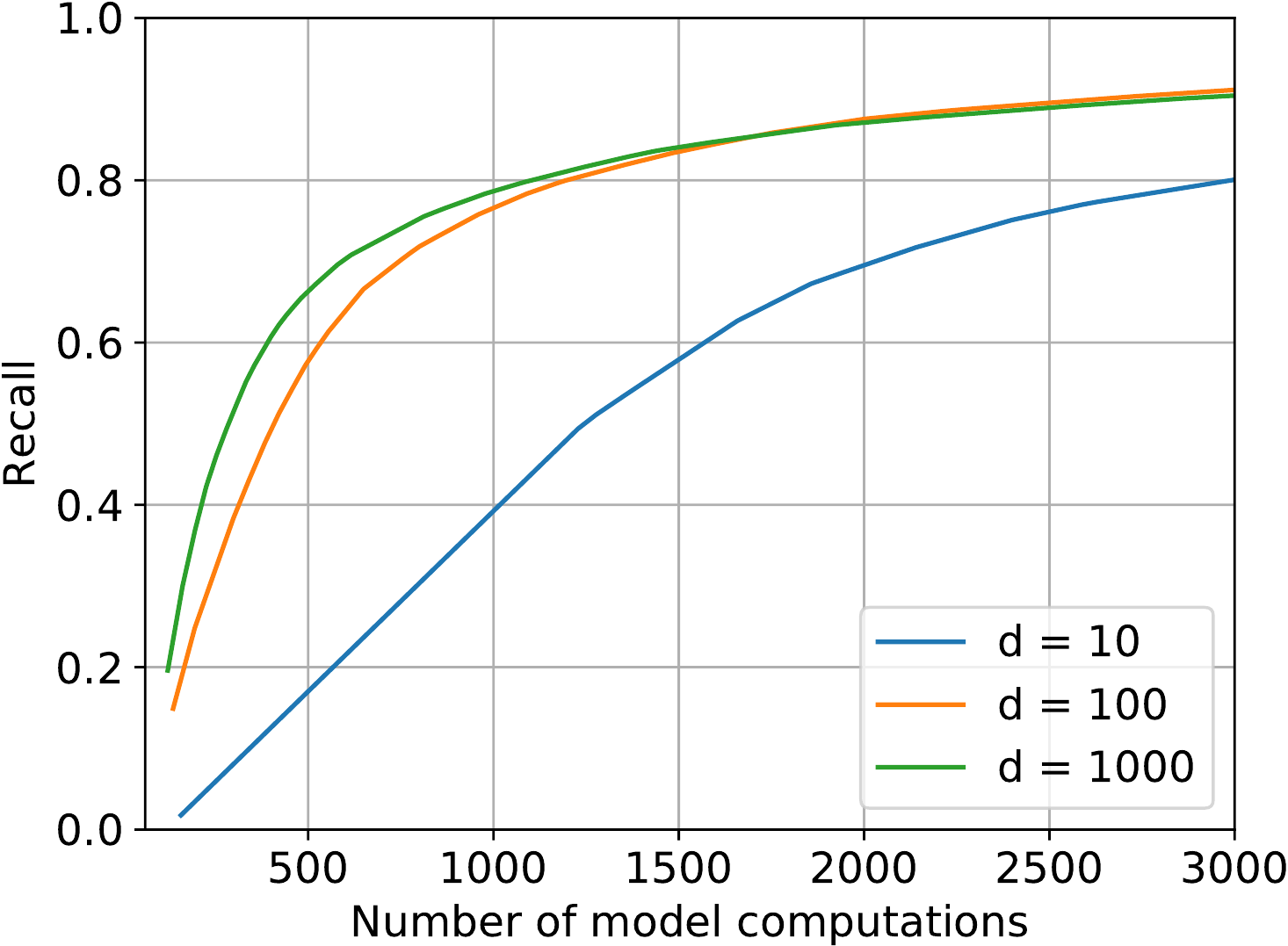}
        \caption{Collections}
        \label{fig:diff_d_recall_collections}
    \end{subfigure}
    \hfill
    \begin{subfigure}[b]{0.326\textwidth}
        \includegraphics[width=\textwidth]{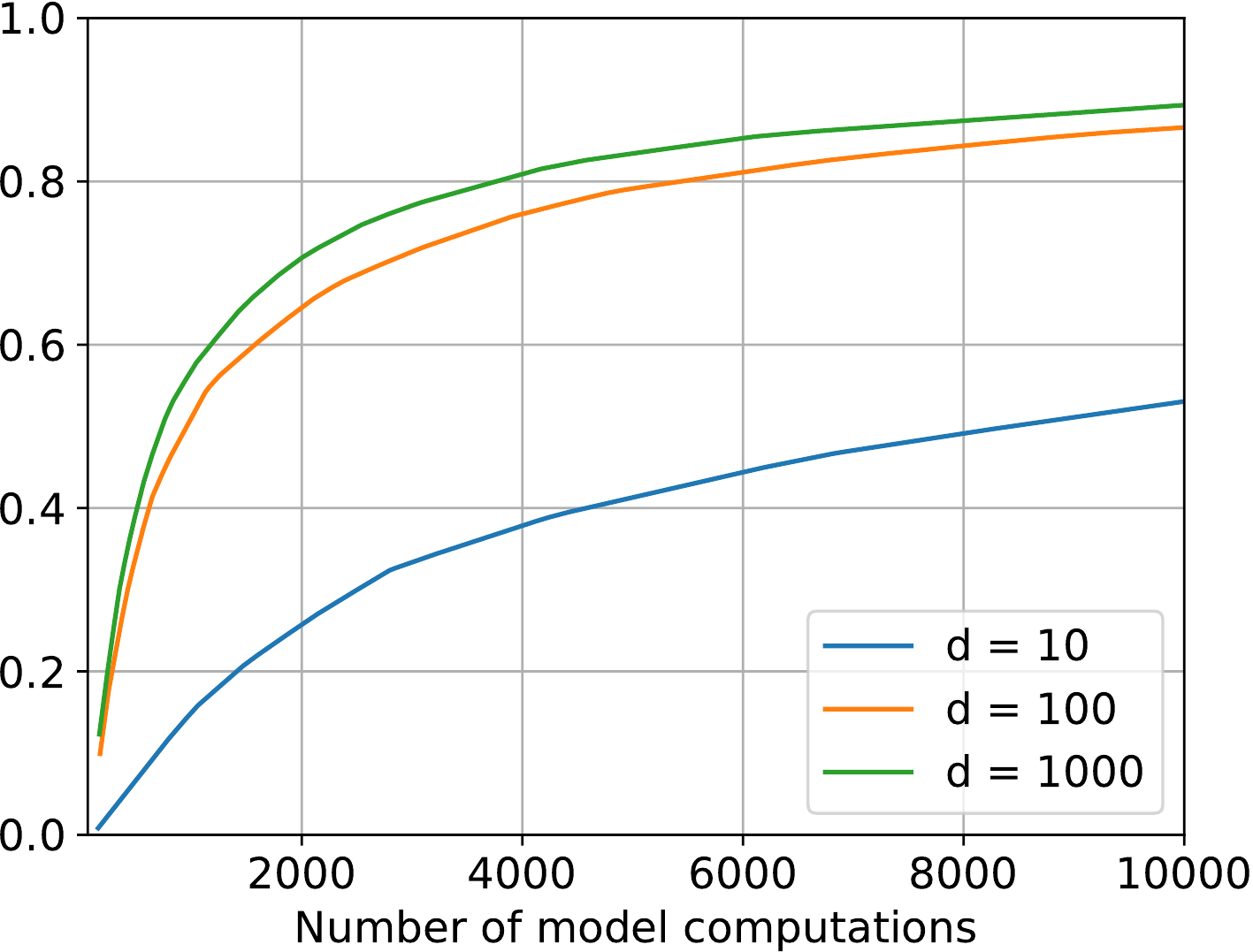}
        \caption{Video}
        \label{fig:diff_d_recall_video}
    \end{subfigure}
    \hfill
    \begin{subfigure}[b]{0.32\textwidth}
        \includegraphics[width=\textwidth]{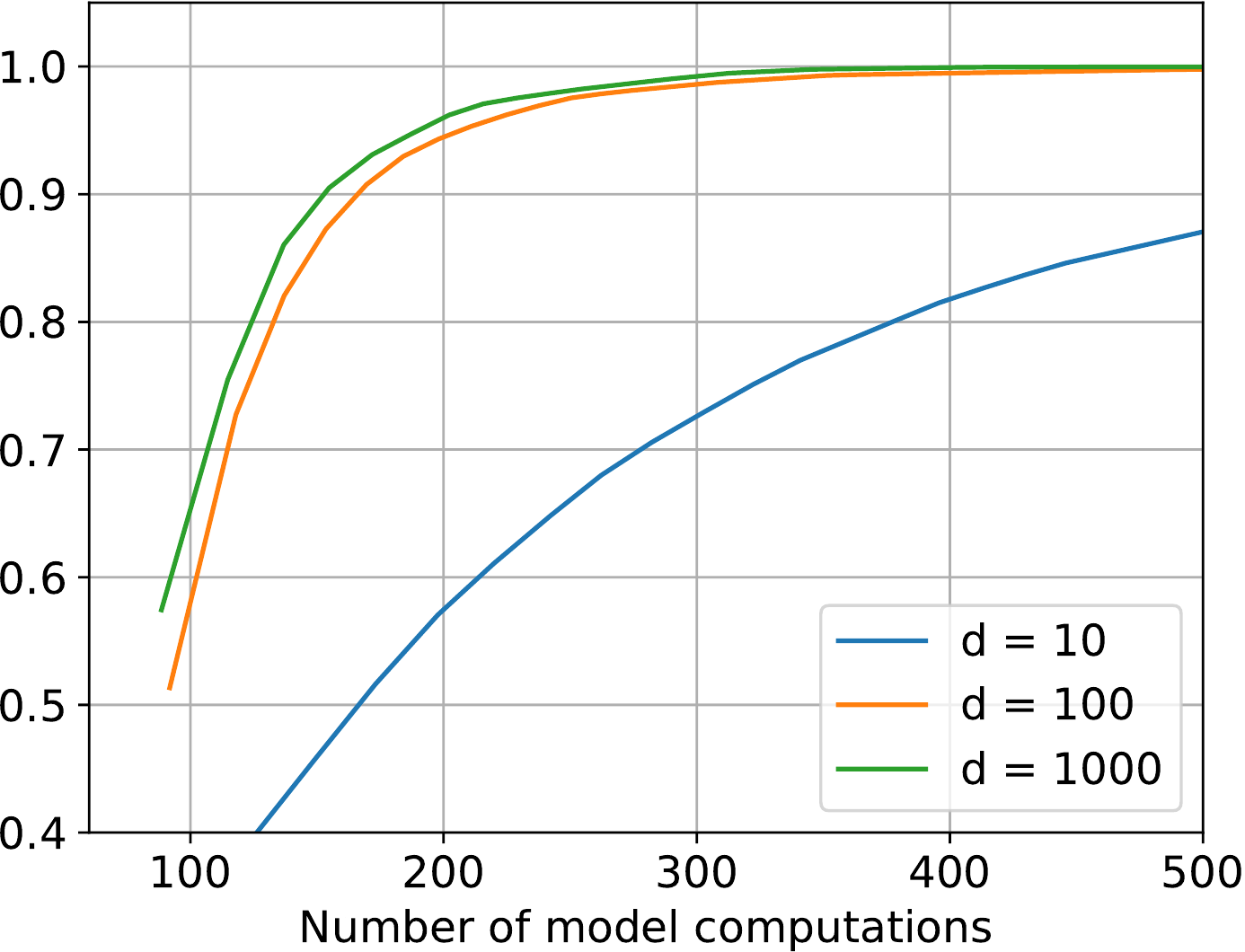}
        \caption{Pinterest}
        \label{fig:diff_d_recall_network}
    \end{subfigure}
    \caption{The dependence on the length of relevance vectors for three datasets in terms of \textit{Recall}. For all datasets we retrieve top-$5$ items.}
    \label{fig:diff_d_recall}
\end{figure*}

\textbf{Length of relevance vectors.} Next, we investigate how the RPG accuracy depends on the length of relevance vectors $d$. We used a random sample of $1,000$ queries for the computation of relevance vectors (\textbf{RPG}) and evaluated recall for $d=10,100,1000$ for all three datasets. The results are shown in Figure~\ref{fig:diff_d_recall} and illustrate \textit{Recall} for different number of model computations. As expected, higher $d$ results in the more accurate retrieval due to the Monte Carlo estimates  in (\ref{precise_similarity}) becoming more accurate. On the other hand, Figure~\ref{fig:diff_d_recall} demonstrates diminishing returns from higher $d$ as the performance difference between $d=100$ and $d=1,000$ is only marginal.

\textbf{Search scalability.} Finally, to investigate the empirical scalability of RPG we varied \textit{Collections} database size $|S|$ in $\{10^3, 10^4, 10^5, 10^6\}$ and determined the number of relevance function computations, required to achieve $0.90$ \textit{Recall} for the top size $K{=}5$. The results, shown in \fig{scalability}, imply the power dependence $|S|^{\alpha}$. Note, however, that the exponent $\alpha$ of the power law is less than $1$ (approximately $1/3$), hence the empirical scalability of RPG is sublinear in database size.

 \begin{figure*}[!ht]
    \centering
    \includegraphics[width=\textwidth]{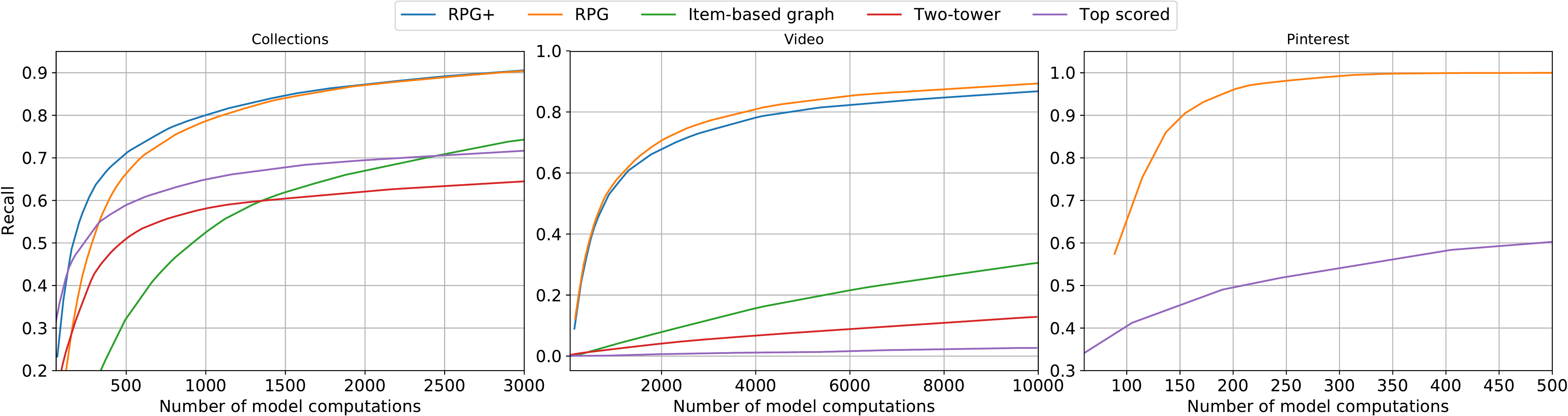}
    \caption{The comparison with baselines for three datasets in terms of \textit{Recall}/number of model computations trade-off. For all datasets we retrieve top-$5$ items.}
    \label{fig:baselines_recall}
\end{figure*}

\begin{figure*}[!ht]
    \centering
    \includegraphics[width=\textwidth]{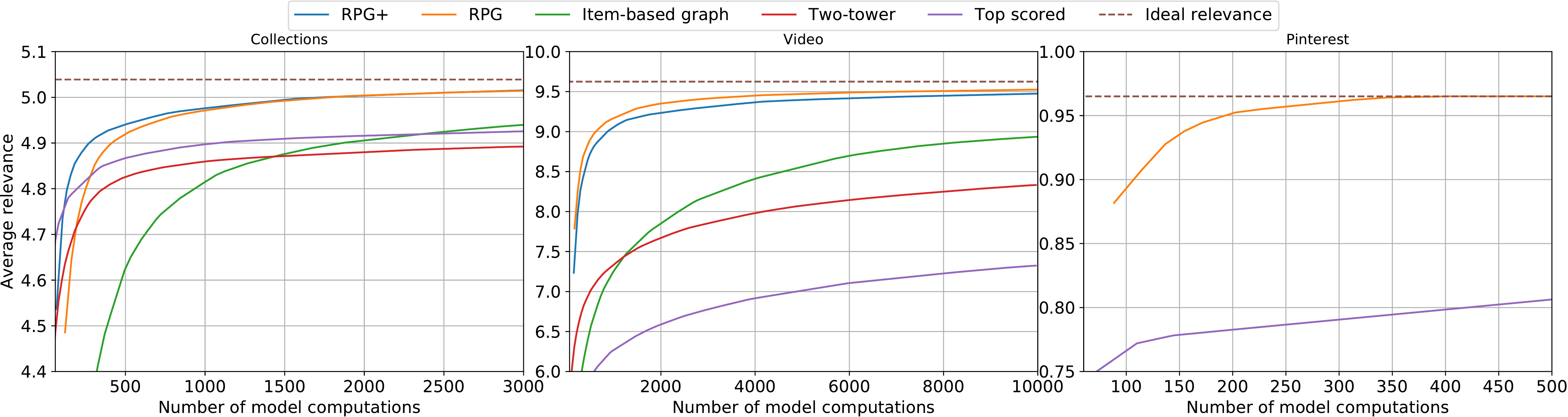}
    \caption{The comparison with baselines for three datasets in terms of \textit{Average relevance}/number of model computations trade-off. For all datasets we retrieve top-$5$ items.}
    \label{fig:baselines_score}
\end{figure*}

 \begin{figure*}[!ht]
    \centering
    \includegraphics[width=\textwidth]{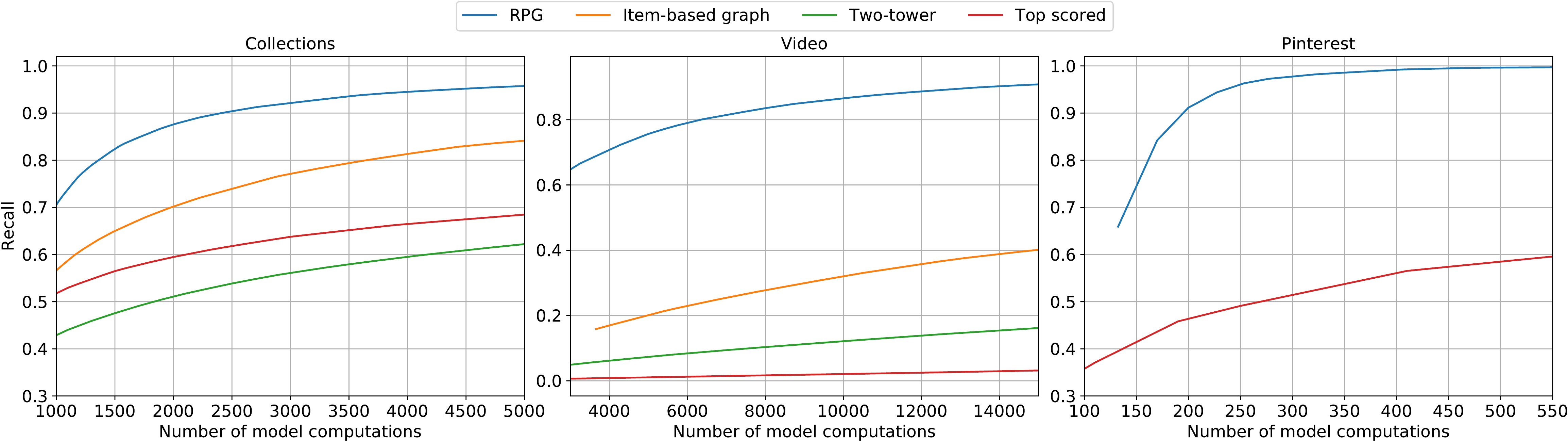}
    \caption{The comparison with baselines for three datasets in terms of \textit{Recall}/number of model computations trade-off. For all datasets we retrieve top-$100$ items.}
    \label{fig:baselines_recall_top100}
\end{figure*}

\subsection{Comparison with baselines}

We compare the proposed RPG method with several baselines for general relevance retrieval. In particular, we evaluate the following methods:

\begin{description}[align=left]
\item [Top-scored] For every item, we compute its average relevance values for train queries and select $N$ items with the maximal global query-independent relevance. Then we rerank these $N$ items based on the actual relevance value, computed by the query-item relevance model. We vary $N$ to achieve runtime/accuracy trade-off. Intuitively, this selects the most "popular" items.  
\end{description}
\begin{description}[align=left]
\item [Item-based graph] is the baseline that uses the similarity graph, constructed on the item features only, instead of the relevance vectors. Let us denote by $h_u$ the $L_2$-normalized vector of features of item $u$. Then the similarity between two items can be defined as
\vspace{0.3cm}
 \begin{equation}
 \label{naivesimilarity}
     s(u, v) = -\| h_u - h_v \|
 \end{equation}
Note, that compared to RPG, the item-based graph has two crucial deficiencies:
 \begin{enumerate}
     \item The item-based graph construction does not use any information about the query distribution or the relevance prediction model.
     \item The dataset could lack item-only features (e.g., \textit{Pinterest}), hence in such cases the item-based graph could not be constructed.
 \end{enumerate}
\end{description}

In practice, one could use a less accurate, computationally cheaper model (e.g., linear) to produce a list of candidates that are then reranked by the expensive GBDT/DNN model. To compare the proposed RPG framework with such two-stage approaches we propose the following baseline:
\begin{description}[align=left]
\item [Two-tower] We learn a "two-tower" DNN that encodes query and item features into 50-dimensional embeddings. The DNN has separate query and item branches, consisting of three fully-connected layers, each having $128$ neurons for \textit{Collections} and $512$ neurons for \textit{Video} with ELU non-linearity and Batch Normalization. The relevance for a query-item pair is predicted as a dot product of the corresponding embeddings. We train this model with the same target as the original GBDT model, with the Adam optimizer \cite{Adam} and OneCycle \cite{OneCycle} learning rate schedule. During the retrieval stage, we select $N$ items that provide maximum dot-product \cite{boytsov2017efficient} with a given query and rerank them based on the actual relevance value. We vary $N$ for runtime/accuracy trade-off. An important weakness of the \textbf{Two-tower} baseline is that it ignores the query-item pairwise features, when producing candidates, and we show that this weakness can be crucial.
\end{description}
Note, however, that the usage of cheaper models for candidate selection could be nicely combined with the RPG search, as described in the following \textbf{RPG+} modification of our approach.
\begin{description}[align=left]
\item [RPG+] The pure \textbf{RPG} uses the same predefined entry vertex to start the graph exploration. However, if there is given a promising candidate from an auxiliary model, then we can use it as an entry point instead. In \textbf{RPG+} we start from the best candidate achieved with the DNN from the \textbf{Two-tower} model. Note, that we do not need any relevance function computations to obtain the candidate. Intuitively, starting from the sufficiently relevant entry vertex, the graph exploration in \textbf{RPG+} requires much smaller hops to reach the "relevant region" of the database.
\end{description}

\begin{table*}
\caption{The importance of different feature groups, computed by the GBDT model.}
\begin{center}
\begin{small}
\begin{sc}
\begin{tabular}{lccc}
\toprule
Dataset & Item features & User features & Pairwise features \\
\midrule
Collections & 0.1466 & 0.0260 & 0.0642\\
Video & 0.0099 & 0.0027 & 0.4114\\
\bottomrule
\end{tabular}
\end{sc}
\end{small}
\end{center}
\label{tab:importances}
\end{table*}

\begin{figure*}
  \centering
  \begin{subfigure}[b]{0.51\textwidth}
    \centering
    \includegraphics[width=90mm]{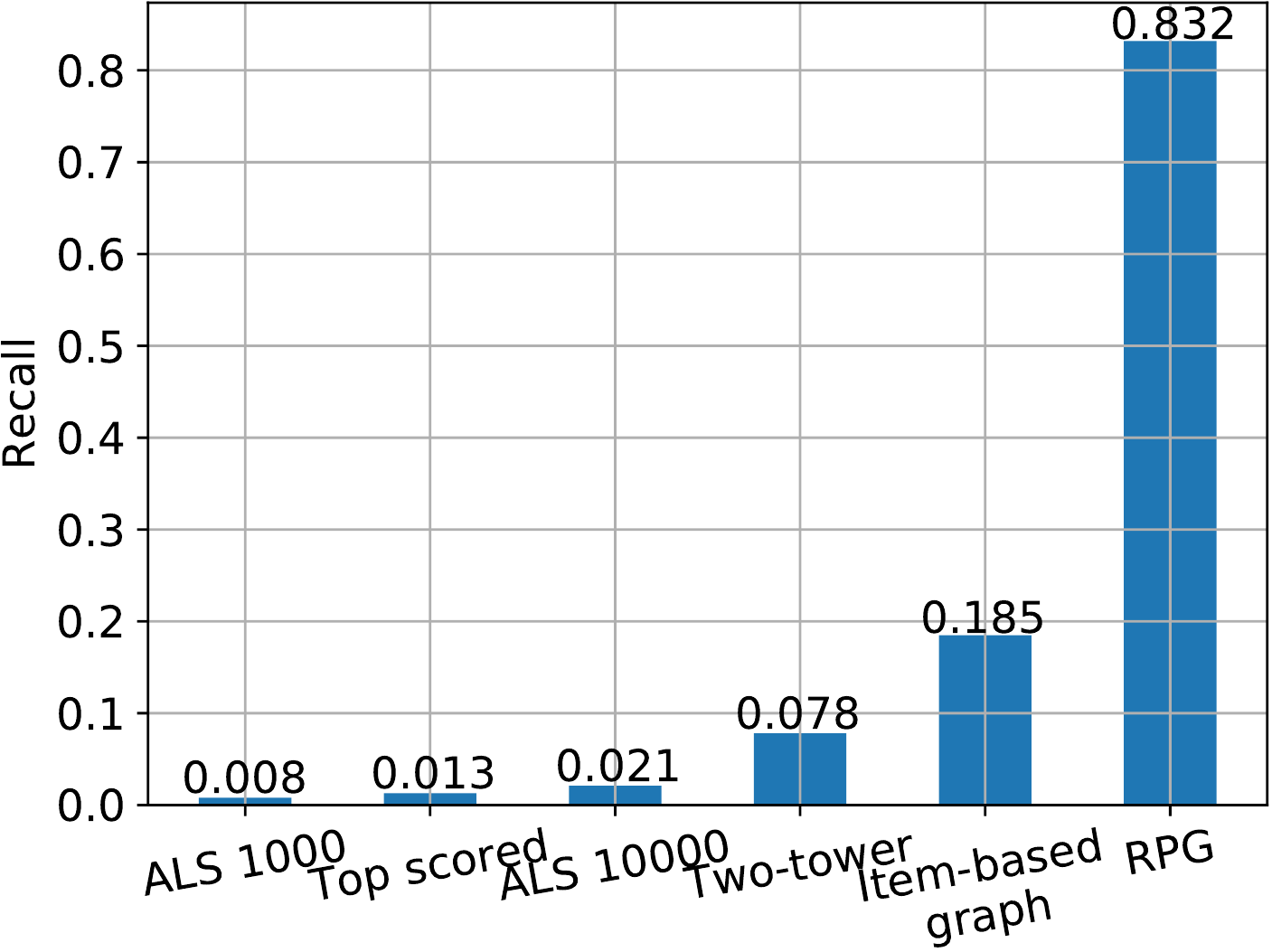}
    \caption{Video}
  \end{subfigure}
  \begin{subfigure}[b]{0.47\textwidth}
    \centering
    \savebox{\tempfig}{\includegraphics[width=90mm]{images/histogram_collections.pdf}}
    \raisebox{\dimexpr\ht\tempfig-\height}{\includegraphics[width=\columnwidth]{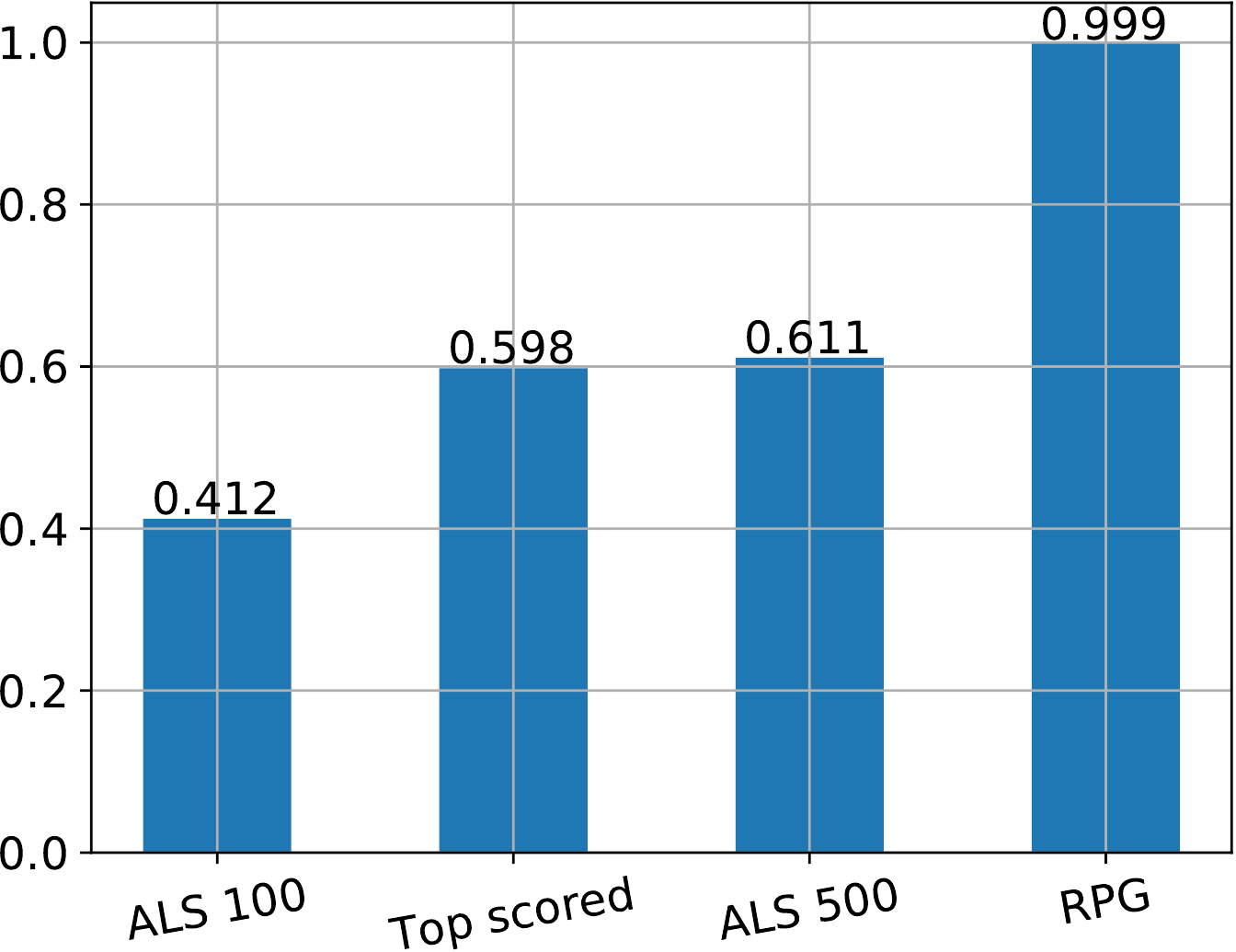}}
    \caption{Pinterest}
  \end{subfigure}\hfill%
  \caption{The comparison of RPG performance with baselines under restriction of $5,000$ relevance function computations for \textit{Video} and $500$ relevance function computations for \textit{Pinterest} dataset.}
  \label{fig:histograms}
\end{figure*}


 Figure~\ref{fig:baselines_recall} and Figure~\ref{fig:baselines_score} present the dependence of \textit{Recall} and \textit{Average relevance} on the number of relevance function computations, respectively. Figure~\ref{fig:baselines_score} reports also the ideal values of \textit{Average relevance}, obtained via exhaustive search. For all datasets, these plots show that the proposed RPG method outperforms all baselines by a large margin in the high-recall regions. Furthermore, \textbf{RPG+} can boost the performance in the low-recall operating point, given a cheap candidate selection model. Note that \textbf{RPG} reaches almost ideal average relevance in a few numbers of model computations. In these experiments, we report the performance when $K{=}5$ items are retrieved, but we claim that \textbf{RPG} consistently outperforms the baselines for larger $K$ as well. Figure~\ref{fig:baselines_recall_top100} presents the dependence of \textit{Recall} on the number of relevance function computations for $K{=}100$ and confirms the superiority of the proposed techniue over baselines.
 
Interestingly, the baselines perform differently on different datasets. In particular, the \textbf{Two-tower} baseline is quite competitive on \textit{Collections}, while giving poor results on \textit{Video}. To explain this observation, we compare the feature importance, computed by the GBDT model\footnote{\url{https://catboost.ai/docs/concepts/fstr.html}}. In a nutshell, for every feature, the importance value shows how the loss function, computed on the train set, changes if this feature is removed. Then we sum the importances across all item, user, and pairwise features and report them in \tab{importances}. Note, that for the \textit{Collections} dataset item features are more important, while for the \textit{Video} dataset the pairwise features contain most signal. Consequently, the \textbf{Top scored} and \textbf{Two-tower} baselines show decent performance on \textit{Collections}, as they could capture the signal from the user and item features and provide precise candidate lists for reranking. Meanwhile, they are not competitive on \textit{Video}, because they lose the information from the pairwise features, which are the most important on this dataset. The \textbf{RPG/RPG+} provides top performance for both datasets.

\subsection{Reducing to matrix factorization problem}

The problem of maximal relevance retrieval can potentially be solved by the matrix factorization methods \cite{mehta2017review}. Let us have a fixed set of queries $P$. Then one can construct embedding vectors $v_i$ for all items and embedding vectors $u_j$ for all the queries that are obtained via a low-rank decomposition of the full relevance matrix $F = V^T U$, where $F_{ij} = f(s_i, p_j)$, $F \in \mathbb{R}^{|S|\times|P|}$, $V \in \mathbb{R}^{r \times|S|}$, $U \in \mathbb{R}^{r\times|P|}$ and $r$ denotes the decomposition rank. Then for a given query we can retrieve $K$ best items in terms of dot product $\langle u, v\rangle$ and then rerank these top-$K$ items exhaustively based on the values of the original relevance function $f(q,v)$. We evaluate the described baseline, performing approximate matrix factorization via Alternating Least Squares implementation from the Implicit library\footnote{\url{https://github.com/benfred/implicit}}. The comparison of ALS with the graph-based methods for two datasets is presented on the \fig{histograms}. On this figure ALS-$N$ means that we randomly selected $N$ items for each query from $P$, computed the corresponding relevance values and performed ALS for the obtained sparse relevance matrix. Note, that the described approach is able to retrieve the relevant items only for queries from $P$ and does not directly generalizes to unseen queries. As operating points, we use $r = 50$ and $K = 5,000$ for \textit{Video} and $r = 20$ and $K = 500$ for \textit{Pinterest}. \fig{histograms} demonstrates that ALS cannot reach the quality of the graph-based methods.
As an upper bound for baselines, which construct dot-product based embeddings for items and users we implemented SVD for matrix $F$. Note, that this is an extremely infeasible baseline as it requires an explicit computation of the full matrix $F$ and this is the same computationally hard as to precompute answers for all the users by exhaustive search. Despite this, SVD still cannot reach the graph methods accuracy. In particular, for the \textit{Video} dataset and $r = 50$, SVD achieves recall $0.693$ and for the \textit{Pinterest} dataset and $r = 20$ it achieves recall $0.988$.


\section{Conclusion}
\label{sect:conclusion}
In this paper, we have proposed and evaluated the Relevance Proximity Graph (RPG) framework for non-exhaustive maximal relevance retrieval with highly-nonlinear models. Our approach generalizes similarity graphs to the scenario, where the relevance function is given for query-item pairs, and there may be no similarity measure for items. Our framework can be applied to a comprehensive class of relevance models, including deep neural networks and gradient boosted decision trees. While being conceptually simple, RPG successfully solves the relevance retrieval problem for million-scale databases and state-of-the-art models, as demonstrated by extensive experiments. As an additional contribution, we open-source the implementation of our method as well as two large-scale relevance retrieval datasets to support further research in this area.

\bibliography{example_paper}
\bibliographystyle{aaai}

\end{document}